\documentclass[prd,aps,twocolumn,a4paper,floatfix,showpacs,footinbib]{revtex4-1}

\usepackage[utf8]{inputenc}  
\usepackage[T1]{fontenc}
\usepackage{graphicx,psfrag}
\usepackage{mathrsfs}
\usepackage{amsmath,amsfonts,amssymb,pifont,gensymb}
\usepackage{multirow,enumerate}
\usepackage{comment,hyperref}
\usepackage{color}
\usepackage{acronym}
\usepackage{xspace}
\usepackage[normalem]{ulem}
\usepackage{mathtools}
\usepackage{subfigure}
\usepackage{makecell}

\newacro{BH}{black hole}
\newacro{NS}{neutron star}
\newacro{PN}{Post-Newtonian}
\newacro{BBH}{binary black hole}
\newacro{BNS}{binary neutron star}
\newacro{EOB}{effective-one-body}
\newacro{NR}{numerical relativity}
\newacro{GW}{gravitational wave}
\newacro{EOS}{equation-of-state}

\newcommand{\be}{\begin{equation}}
\newcommand{\ee}{\end{equation}}
\newcommand{\bea}{\begin{eqnarray}}
\newcommand{\eea}{\end{eqnarray}}
\newcommand{\bel}{\begin{align}}
\newcommand{\eel}{\end{align}}

\newcommand{\linf}{\texttt{LALInference}}

\def\Msun{{\rm M_{\odot}}}
\def\GMc2{{\rm G M_{\odot} c^{-2}}}
\def\Mpc{{\rm Mpc}}

\def\SEOBNRv4T{\texttt{SEOBNRv4T}\xspace}

\def\Msun{\rm M_\odot}

\usepackage{color}
\definecolor{cyan}{rgb}{0,0.9,0.9}
\definecolor{orange}{rgb}{0.9,0.5,0}
\definecolor{magenta}{rgb}{1,0,1}
\definecolor{purple}{rgb}{0.8,0.4,0.8}
\definecolor{gray}{rgb}{0.5,0.5,0.5}
\definecolor{mygreen}{rgb}{0.1,0.8,0.1}
\definecolor{darkblue}{rgb}{0.0,0.0,0.6}

\begin{document}

\title{Constructing Love-Q-Relations with Gravitational Wave Detections} 

\author{Anuradha Samajdar$^{1,2}$}
\author{Tim \surname{Dietrich}$^{1,3}$}

\affiliation{${}^1$ Nikhef, Science Park 105, 1098 XG Amsterdam, The Netherlands}
\affiliation{${}^2$ Department of Physics, Utrecht University, Princetonplein 1, 3584 CC Utrecht, The Netherlands}
\affiliation{${}^3$ Institut f\"{u}r Physik und Astronomie, Universit\"{a}t Potsdam, Haus 28, Karl-Liebknecht-Str. 24/25, 14476, Potsdam, Germany}

\date{\today}

\begin{abstract}
Quasi-universal relations connecting the tidal deformability and 
the quadrupole moment of individual neutron stars are predicted by theoretical computations, 
but have not been measured experimentally. 
However, such relations are employed during the interpretation of gravitational waves
and, therefore, have a direct impact on the interpretation of real data. 
In this work, we study how quasi-universal relations can be tested and measured 
from gravitational wave signals connected to binary neutron star coalescences. 
We study a population of $120$ binary neutron star systems and find that Advanced LIGO and 
Advanced Virgo at design sensitivity could find possible deviations of predicted 
relations if the observed neutron stars are highly spinning. 
In the future, a network of third generation (3G) detectors 
will be able to even allow a measurement of quasi-universal relations. 
Thus, the outlined approach provides a new test of 
general relativity and nuclear physics predictions.
\end{abstract}

\maketitle

\section{Introduction}
\label{sec:intro}

The observation of GW170817 proved that gravitational waves (GWs) serve as a new observational window 
to probe matter at supranuclear densities and to decode the unknown equation of state 
(EOS) governing the neutron star's interior~\cite{TheLIGOScientific:2017qsa,Abbott:2018exr,Abbott:2018wiz}. 
Already from this single detection, 
it was possible to place constraints on the supranuclear EOS,
e.g.,~\cite{TheLIGOScientific:2017qsa,Annala:2017llu,
Capano:2019eae,Bauswein:2017vtn,De:2018uhw,Margalit:2017dij,
Abbott:2018exr,Most:2018hfd,Coughlin:2018miv,Radice:2018ozg} 
and to disfavor some of the theoretical predictions. 
The recent detection of another binary neutron star (BNS) merger, GW190425~\cite{Abbott:2020uma} 
however, does not shed additional light on EOS information because of its high mass~\cite{Coughlin:2019xfb,Coughlin:2019zqi}. 
Nevertheless, rate estimates for BNS coalescences 
($250-2810 \ \rm Gpc^{-3} yr^{-1}$~\cite{Abbott:2020uma}) 
show that we can expect many more BNS signals 
to be detected in the near future.

During a BNS coalescence, each neutron star undergoes tidal deformation due to the influence of 
the other star's gravitational field. This tidal deformability is imprinted in the emitted GW signal and 
carries information about the internal structure of the star. The main quantity characterizing these tidal deformations is 
the tidal polarizability $\Lambda= 2k_2/(3 C^5)$ with $k_2$ being the tidal Love number describing 
the static quadrupolar deformation of one neutron star in the gravitoelectric field of the companion and 
$C$ being the compactnesses of the star at isolation. 

In addition, a spinning neutron star undergoes deformation, 
% even without an external gravitational field. 
encoded in an additional \emph{spin-induced} 
quadrupole moment. 
For rotating neutron stars, the quadrupole moments vary as 
$\mathcal{Q} \simeq -Q \chi^2 m^3$ with $\chi$ and $m$ being the dimensionless spin and the
mass of the object; see~\cite{Laarakkers:1997hb} for a first discussion and 
Ref.~\cite{Pappas:2012ns} for an upgrade and update of~\cite{Laarakkers:1997hb}. 
Here, $Q$ is a parameter connected to the 
internal structure of the neutron star depending on the supranuclear EOS. 
For a given EOS, this relation may be written as $\mathcal{Q} \simeq -Q(m)\chi^2$. 
The corresponding imprint in the GW phasing from $\mathcal{Q}$ was computed in~\cite{Poisson:1997ha}. 
Refs.~\cite{Harry:2018hke,Samajdar:2019ulq} laid out the importance of the quadrupole moment on the 
measurability of parameters in GW signals for highly spinning NSs and \cite{Agathos:2015uaa} 
investigated possible effects on GW signals introduced by the spin-induced quadrupole moments by combining 
information from multiple signals.
% investigated possible biases introduced by the spin-induced quadrupole moment 
% once information from multiple signals are combined.
Finally, \cite{Krishnendu:2017shb} used the measurement of spin-induced quadrupole moments as a probe to distinguish 
between a binary black hole signal within general relativity  
and a signal arising from a binary of exotic compact objects. 
The analysis was further extended to a Bayesian approach in~\cite{Krishnendu:2019tjp},  
the only work which samples directly on the spin-induced quadrupole moment parameters. \\

Most analyses performed on GW signals GW170817 and GW190425 inferred the 
quadrupole moment of each neutron star from their tidal deformabilities, by leaving 
the latter as free parameters and using the EOS-insensitive relations~\cite{Abbott:2018exr,Abbott:2018wiz,Abbott:2020uma} 
to determine the spin-induced quadrupole moment.
Quasi-universal relations connecting the tidal deformability and the spin-induced 
quadrupole moment of neutron stars have been first introduced by Yagi and Yunes~\cite{Yagi:2013bca}
and have been improved by incorporating information from GW170817~\cite{Carson:2019rjx}. 
While these EOS-insensitive relations are to second order in the slow-rotation approximation 
essentially independent of the NS spin, additional deviations may occur for fast rotating neutron stars~\cite{Yagi:2013bca}.
However, we point out that these relations have been employed for the analysis of GW170817 and GW190425 even beyond the neutron star's breakup spin. 
Therefore, we want to ask the question whether it is possible to verify and potentially measure the relation between the 
quadrupole moment and the tidal deformability from real GW data. 
For this purpose, we use the Yagi-Yunes relation~\cite{Yagi:2013bca,Yagi:2013awa} that connects the quadrupole moment to the tidal
deformability:
\begin{equation}
 \ln{Q} = a_i + b_i\ln{\Lambda} + c_i\ln{\Lambda}^2 + d_i\ln{\Lambda}^3 + e_i\ln{\Lambda}^4, 
 \label{eqn:qu}
\end{equation}
with the fitting parameters $a_i = 0.194$, $b_i=0.0936$, $c_i=0.0474$, $d_i=-4.21\times10^{-3}$, and $e_i=1.23\times10^{-4}$.
We use the quadrupole moments of the individual stars as 
free parameters and sample on them during the analysis instead of relying on the existing quasi-universal 
relations to infer them from their corresponding tidal deformability parameters. 
While this increases the dimensionality of the problem and leads to larger uncertainties in the 
observed parameters, it also allows to test and find relations between the quadrupole moment and
the tidal deformability. 

\section{Methods}
\label{sec:method}

We perform a Bayesian analysis for parameter estimation using the 
\linf~module~\cite{Veitch:2014wba} available in the \texttt{LALSuite}~\cite{lalsuite} package. 
We employ the nested sampling algorithm to estimate 
posterior probability distribution functions~\cite{Veitch:2009hd, skilling2006} 
which further encode information about the parameters.
The parameter set of a BNS source consists of  
$\{ m_1, m_2, \chi_1, \chi_2, \theta, \phi, \iota, \psi, D_L, t_c, \varphi_c, 
\Lambda_1, \Lambda_2\}$. $m_i$ is the mass of the $i^{\rm th}$ object, 
$\chi_{i} = \frac{\vec{S}_i}{m_i^2}\cdot \hat{L}$ is the dimensionless
spin parameter aligned with the direction of the orbital angular
momentum $\hat{L}$, $\theta$ and $\phi$ are the angular coordinates denoting the sky location, 
$\iota$ and $\psi$ are the angles describing the binary's orientation with respect to the line of sight, 
$D_L$ is the luminosity distance to the source, $t_c$ and $\varphi_c$ are the time and phase 
at the instance of coalescence, and $\Lambda_i$ are the dimensionless tidal deformability parameters. 
In addition, our parameter set also includes the spin induced quadrupole moments $dQ_1=Q_1-1$ 
and $dQ_2=Q_2-1$. 

For our simulations, we employ the aligned spin waveform model 
\texttt{IMRPhenomD\_NRTidalv2}~\cite{Dietrich:2019kaq}. 
Unlike in~\cite{Dietrich:2019kaq}, our model contains amplitude tidal corrections and 
higher-order spin-squared and spin-cubed terms at 3.5PN along with their 
corresponding spin-induced quadrupole moments, in addition to the spin-induced 
quadrupole moment terms at 2PN and 3PN. 
We simulate $120$ sources in random noise realizations. 
The component masses lie between 1.0$\Msun$ and 2.0$\Msun$. Their 
tidal deformabilities are computed assuming the ALF2 EOS~\cite{Alford:2004pf}, 
which is a hybrid EOS with the variational-method APR EOS for nuclear matter~\cite{Akmal:1998cf} 
transitioning to color-flavor-locked quark matter. ALF2 has been picked since it is in agreement with 
recent multi-messenger constraints on the EOS~\cite{Coughlin:2018miv}.
The sources are distributed uniformly in co-moving volume between 
15$\Mpc$ and 150$\Mpc$ with randomly chosen inclination angles and 
random sky locations. The dimensionless spin components 
are distributed uniformly between $-0.5$ and $0.5$, 
while these values are significantly larger than observed in BNS systems, 
neutron stars not bound in BNS systems can rotate very rapidly, e.g., 
PSR J1807$-$2500B with a rotation frequency 
of $239$Hz~\cite{Lorimer:2008se,Lattimer:2012nd}. 
Furthermore, the recent observation of GW190425~\cite{Abbott:2020uma} 
whose estimated individual masses are inconsistent with the population of 
observed galactic BNSs showed that an extrapolation 
from our limited number of known galactic BNS systems is unreliable so that we include 
also higher spins in our investigation. 
We consider two injection sets for our simulated sources; (i) one where the injected 
quadrupole-monopole moments computed from the quasi-universal relation in Eq.~\eqref{eqn:qu}, i.e., 
$Q_{\rm injection} = Q_{\rm Yagi-Yunes}$, 
and (ii) one where the injected quadrupole-monopole moments do not 
follow the quasi-universal relation; the injected moments here are half the values 
computed from Eq.~\eqref{eqn:qu} as an arbitrary choice of a modified quasi-universal relation, i.e., 
$Q_{\rm injection} = 1/2 \times Q_{\rm Yagi-Yunes}$.
Modified relations may occur in alternate theories of gravity like the dynamical Chern Simons 
theory~\cite{Alexander:2009tp}; cf. e.g.~\cite{Yagi:2013awa}. 
In both kinds of injections, the quadrupole moments $dQ_i$ 
are sampled uniformly between $[0,30]$ and the tidal deformabilities $\Lambda_i$ are sampled uniformly between $[0,5000]$. 
As for the other parameters, we 
sample the chirp mass uniformly between 0.7 $\Msun$ and 2 $\Msun$, the mass ratio $m_2/m_1$ 
is sampled uniformly between 1/8 and 1, and 
the spin components are sampled uniformly between $[-0.7,0.7]$.
% , i.e., 
% we allow for a larger spin in the recovery than during the injection.

\section{Results}
\label{sec:results}

\begin{figure}[t]
\includegraphics[width=0.48\textwidth]{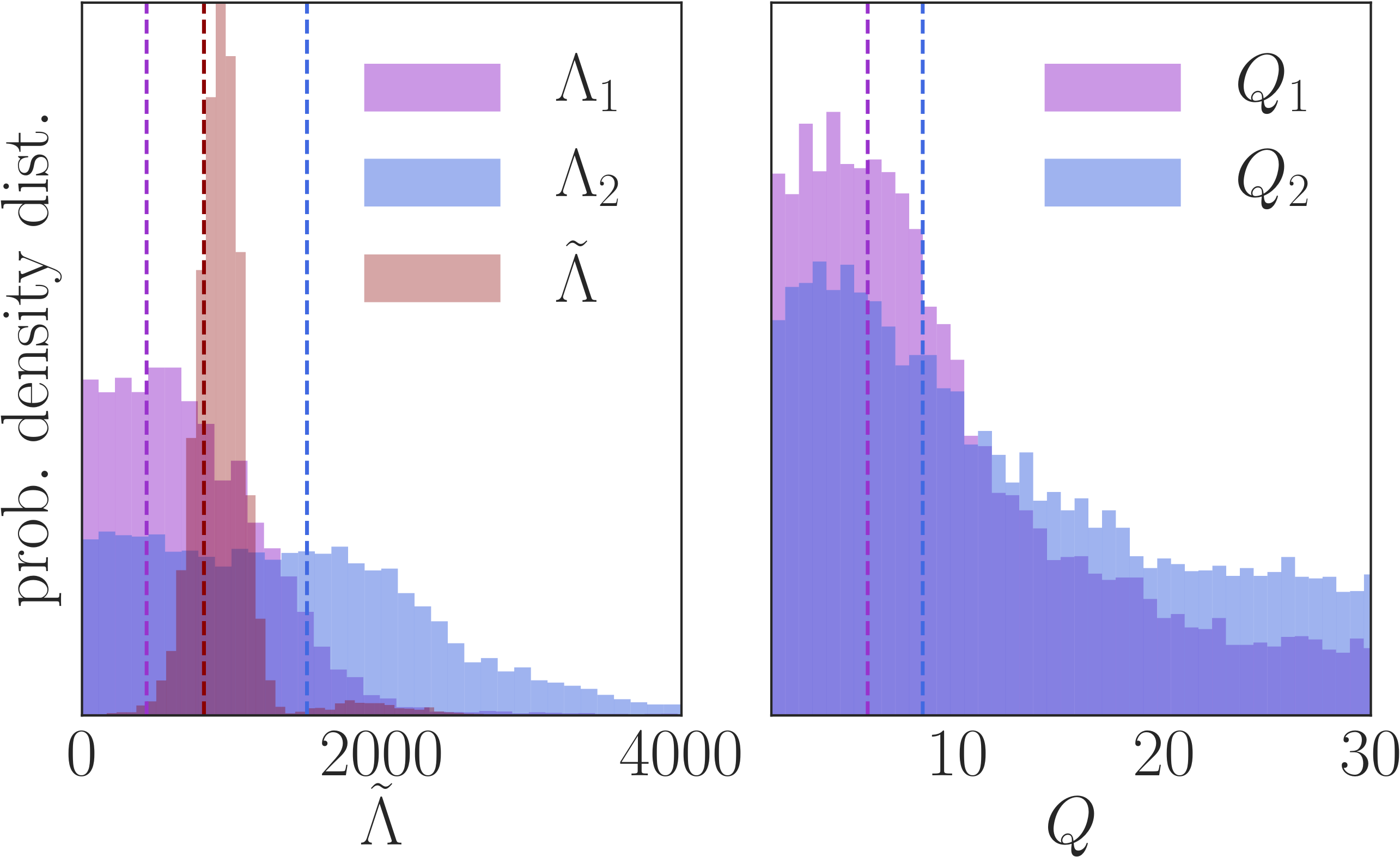}
\caption{Posterior probability distributions 
of $\Lambda_1,\Lambda_2,\tilde{\Lambda},Q_1,Q_2$
from our set of injections. This particular setup has a signal-to-noise-ratio of 33.45. 
The neutron star masses are $m_1=1.472653, m_2=1.185832$, 
the dimensionless spins are $\chi_1=0.496,\chi_2=-0.072$. 
Employing the ALF2 EOS, the tidal deformabilities are $\Lambda_1=431,\Lambda_2=1501$.
The injected values are shown as vertical dashed lines. 
In particular due to the large spin of the primary object, 
this setup is one of the few cases for which the individual tidal 
deformabilities and quadrupole moments can be determined with 
the advanced LIGO and advanced Virgo network.}
\label{fig:example}
\end{figure}
%Example for the recovery
\begin{figure}[t]
\includegraphics[width=0.48\textwidth]{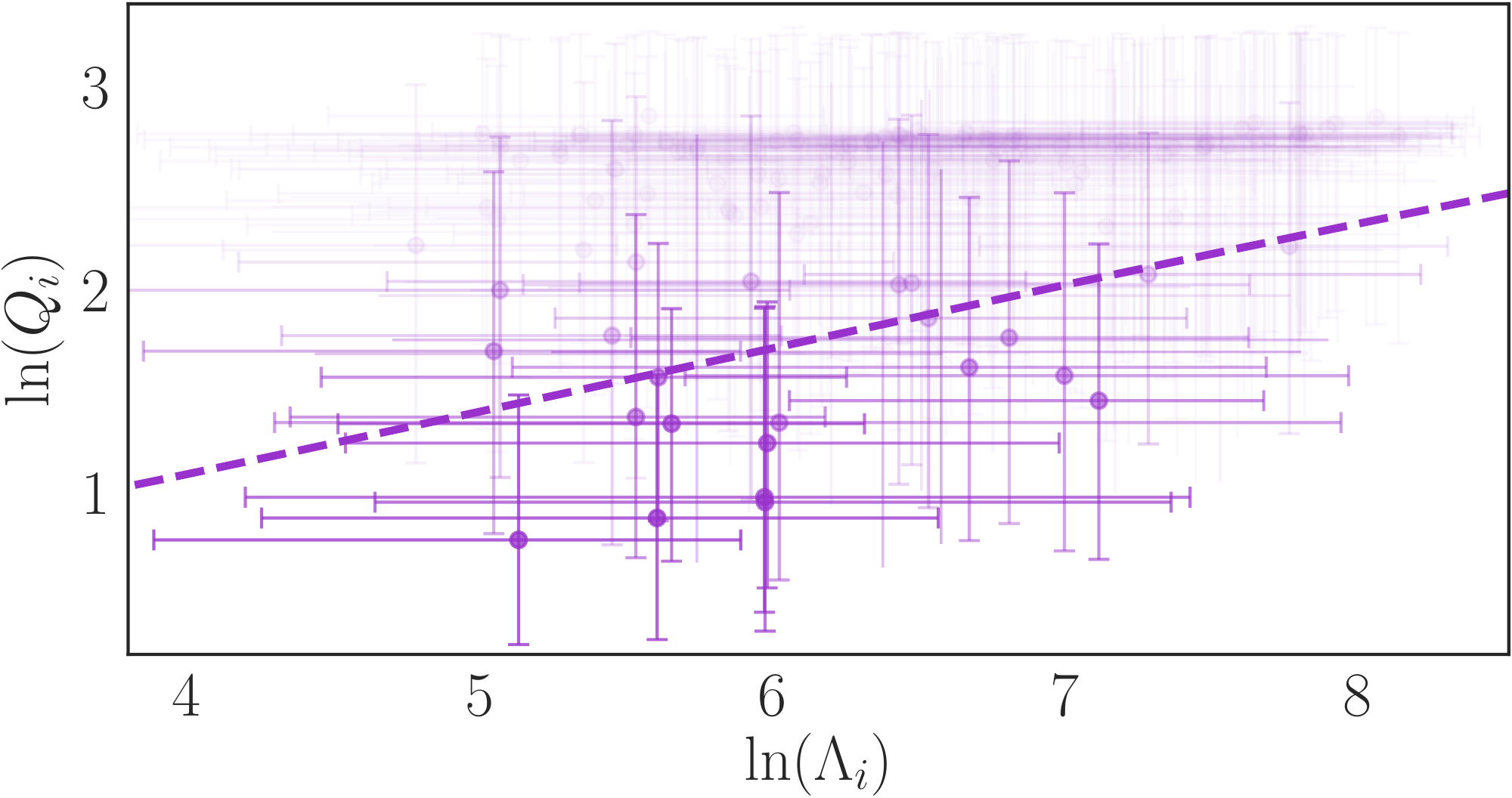}
\includegraphics[width=0.48\textwidth]{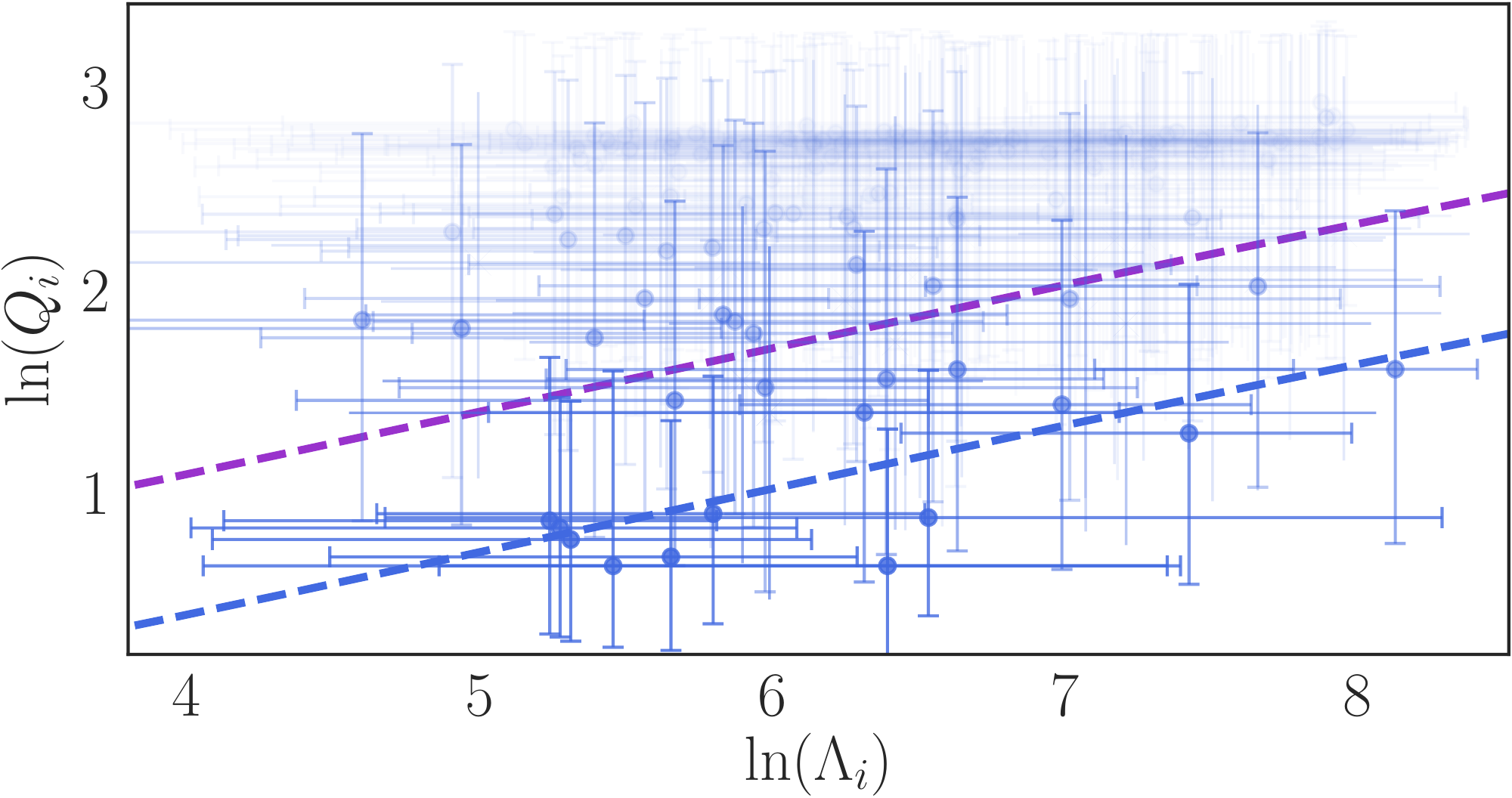}
\caption{Recovered $\Lambda_{1,2}$ and $Q_{1,2}$ values from our simulated population of 120 
BNS systems for a 2G detector network. The shown errorbars mark the $1\sigma$-credible interval, while the individual markers 
refer to the $50\%$-percentile. 
Fainther crosses refer to data with larger uncertainties. \\
Top panel: The injection set based on the quasi-universal relation Eq.~\eqref{eqn:qu}.
The dashed line refers to the quasi-universal relation 
predicted by Yagi and Yunes, Eq.~\eqref{eqn:qu}. \\
Bottom panel: The dashed purple line refers to the Yagi-Yunes quasi-universal relation , 
the blue dashed line the modified relation where $Q$ is reduced by $50\%$ with respect to 
Eq.~\eqref{eqn:qu}, i.e., to the values used for the injection set. }
\label{fig:QM_2G}
\end{figure}

\paragraph*{\textbf{Testing existing quasi-universal relations:}}

Based on the methods discussed before, 
we extract from our simulated BNS population the individual tidal deformabilities of the two stars 
($\Lambda_1$ and $\Lambda_2$) and the spin-induced quadrupole moments $Q_1$ and $Q_2$. 
As an example, we show the recovery of one injection in Fig.~\ref{fig:example}. 
For the shown example, the injected values of $Q_{1,2}$ are 
determined from Eq.~\eqref{eqn:qu}, i.e., we assume the correctness of the theoretically 
derived quasi-universal relations for the injection. 
We find that the vast majority of detections will not allow us to determine reliably 
the individual parameters $\Lambda_{1,2},Q_{1,2}$. 
This is understandable since the individual parameters enter in the GW phase description 
in special combinations, e.g., tidal effects are dominated by the 
tidal deformability parameter 
\begin{equation}
\tilde{\Lambda}=\frac{16}{13} \sum_{i=1,2} \Lambda_i \frac{m_i^4}{M^4}\left( 12-11\frac{m_i}{M} \right),  
\label{eq:Lambda_tilde}
\end{equation}
see e.g.~\cite{Flanagan:2007ix} and Fig.~\ref{fig:example} for an illustration.
Unfortunately, for the interpretation of quasi-universal relations for single neutron stars, 
we have to measure accurately the parameters of the individual stars. 

In Fig.\ref{fig:QM_2G} (top panel) we show all 240 recovered values, 
for both components of $Q_i$ and $\Lambda_i$, 
together with their $1\sigma$-credible interval, 
where we point out that in particular the lower bound on the tidal deformabilities and quadrupole
moments are partially driven by the choice of our prior, i.e., that $\Lambda_i\geq0$ and $Q\geq1$.
Simulations whose individual parameters return the prior are shown as faded. 
Only a few simulations have large enough signal-to-noise ratios as well as high
individual spins 
so that about 15 out of a total of 240 individual parameters can be measured reliably. 
Among these sources, the lowest component spin is $\sim 0.2$. 
In almost all of these cases, these parameters belong to the more massive star in the 
binary system since its tidal deformability and spin-induced quadrupole moment dominate. 

For all systems for which $Q_i$ can be measured, the predicted quasi-universal relation 
connecting $Q-\Lambda$ lies within the $1\sigma$-credible interval, which shows that, 
in principle, an assessment of the robustness of Eq.~\eqref{eqn:qu} is possible.\\

\paragraph*{\textbf{Probing new $\Lambda$-$Q$ relations:}}

To answer the question if we would be able to detect a violation of Eq.~\eqref{eqn:qu}, 
we have analysed the same set of injections, i.e., identical parameters except for a reduction 
of the quadrupole moments $Q_{1,2}$ by $50\%$. 
We show the recovered parameters in Fig.~\ref{fig:QM_2G} (bottom panel). 
As before, most of the simulations do not allow a reliable extraction of the quadrupole moments and the 
individual tidal deformabilities, however, for highly spinning and close systems, 
we find a set of data which are not in agreement with the Eq.~\eqref{eqn:qu} (purple line), 
but with the modified relation for which $Q_{\rm new} = Q/2$. 
Obviously, the particular choice of $Q_{\rm new}$ is arbitrary, 
however, it shows that large enough deviations from existing theoretical 
predictions might already be measurable with the 
second generation (2G) GW detectors~\footnote{We expect that this observation does 
not depend on the particular EOS, but that only the deviation from Eq.~\eqref{eqn:qu} is important}. \\

\begin{figure}[t]
\includegraphics[width=0.48\textwidth]{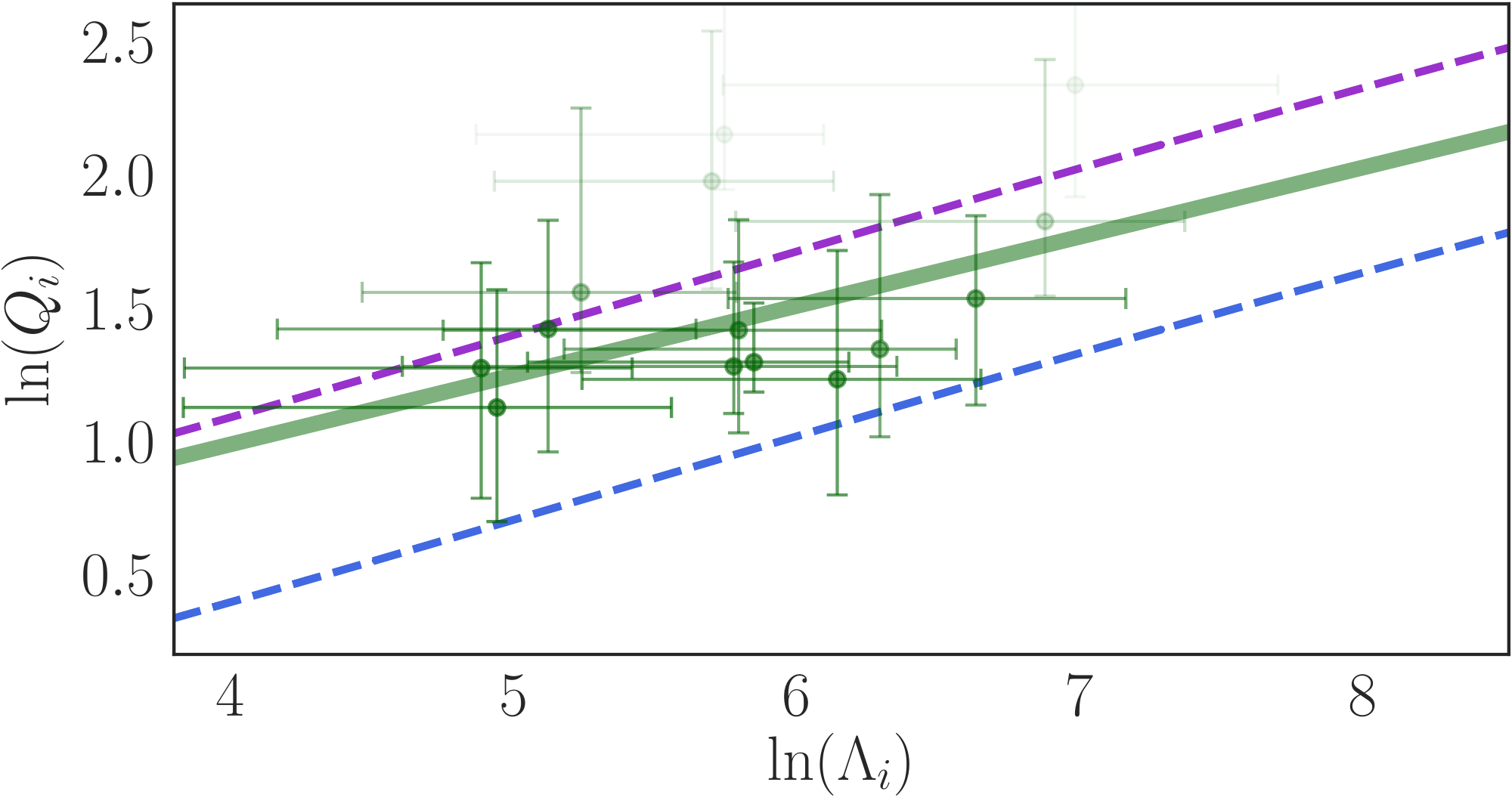}
\includegraphics[width=0.48\textwidth]{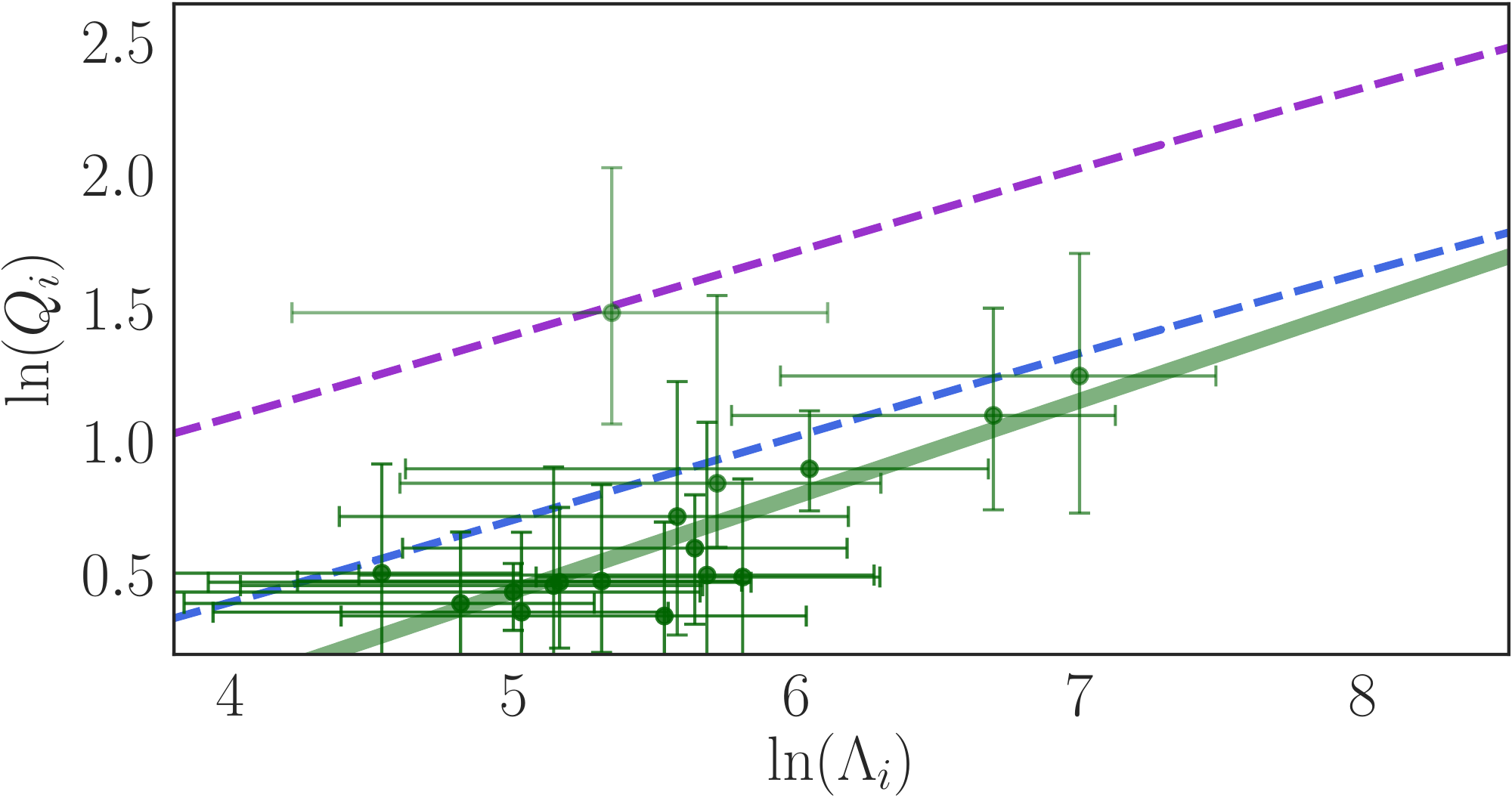}
\caption{Recovered $\Lambda_{1,2}$ and $Q_{1,2}$ values from our simulated population of 120 
BNS systems for which $\log(Q_i)>2.5$ and $\Delta \log(Q)<1$ for a 3G detector network. 
The shown errorbars ($\Delta \log \Lambda$) mark the $1\sigma$-credible interval. 
Fainther crosses refer to data with larger uncertainties.
The dashed purple line refers to Eq.~\eqref{eqn:qu}, the blue dashed line to the quasi-universal 
for which $Q_i$ got reduced by $50\%$, and the green solid line refers to the best fit of the data.\\
Top panel: The injection set is based on Eq.~\eqref{eqn:qu}.
Bottom panel: The injection set is based on our modified quasi-universal relation.}
\label{fig:QM_3G}
\end{figure}

\paragraph*{\textbf{Construction of $\Lambda$-$Q$ relations with 3G detectors:}}
Finally, we simulate these sources in noise generated with envisaged sensitivities 
of future third generation (3G) detectors. For the 3G detectors, we use the noise 
curve of the Einstein Telescope (ET) detector with its ET-D configuration~\cite{Hild:2010id}, 
a cryogenic detector to be built underground within the 
next decade in Europe~\cite{Punturo_2010}, referred to as `ET'. 
Ref.~\cite{Dwyer:2014fpa} introduced the idea of an interferometer available within similar timelines in the USA, 
also known as `Cosmic Explorer' (CE). Unlike ET, CE is planned to be a ground-based detector with an arm length 
of 40 km. For our configuration, we choose a detector network including the ET detector, with its 
xylophone configuration (located at the Virgo site) and two CE-type detectors (located at the two LIGO 
sites)~\footnote{We note that we include an older CE noise curve and that we do not take a 
frequency-dependent response into account, we expect that due to our relatively large initial frequency, 
the effect of the frequency-dependence of the response function is small. 
We refer the reader to~\cite{Chen:2020fzm} for additional details.}. 
The 3G detectors will have the ability to reach lower cutoff frequencies of $f_{\rm low} \sim 1$ Hz, 
which means that sources like those considered before, i.e., for 2G detector network, 
will spend many more cycles in the 3G detectors' band, therefore improving 
both the signal-to-noise ratio as well as the duration for which the signal is visible in band.
Due to limited computational resources, we keep the lower frequency cutoff with the 3G detectors 
same as the simulations with design sensitivity of advanced LIGO and advanced Virgo, i.e., 
$f_{\rm low} =28$ Hz. 
While this means that we are not using 
the full potential of the future detectors and that we artificially reduce the maximum SNR~\footnote{
Although we use a $f_{\rm low} =28$ Hz for the 3G network, we obtain SNR values of about $10^2$ up to $10^3$, 
i.e., about $20$ to $30$ times larger than for the 2G network.}, 
this procedure leads to a conservative result, i.e., 
the result will be better with future data-analysis techniques. \\

Employing the 3G network described above, we present the extracted values of $\Lambda_i$ and $Q_i$
for our two injection sets in Fig.~\ref{fig:QM_3G}, where we restrict to using the data for which 
(i) $\log(Q)\leq2.5$, larger values are basically not expected and 
an indicator that the prior is recovered and (ii) we remove all datapoints for which 
$\Delta \log(Q) > 1$, where $\Delta \log(Q)$ refers to the width of the $1\sigma$ credible interval 
in the log-log plot, Fig.~\ref{fig:QM_3G}.
We find clearly that the recovered source parameters cluster around the respective, injected 
quasi-universal relations. 

For a quantitative measure, we try to extract a phenomenological $Q-\Lambda$ 
relations directly from our recovered dataset. 
We fit the datapoints shown in Fig.~\ref{fig:QM_3G} according to 
\begin{equation}
 \ln{Q} = \hat{a}_i + \hat{b}_i\ln{\Lambda}.  
 \label{eqn:qu_fit}
\end{equation}
For the fitting, we use weights that are indirectly proportional to the size of the $1\sigma$ 
credible interval of $Q_i$, i.e., setups in which the induced quadrupole-moment is 
measured more accurately are favored. 
Different to Eq.~\eqref{eqn:qu} we decided to remove higher order terms 
since the measurement uncertainties do not allow any reliable 
determination of terms $\propto \log(\Lambda)^k$ with $k>1$. 
We find $\hat{a} = - 0.05014 , \ \hat{b}= 0.2595$ for the dataset shown in the top panel, i.e., 
those simulations employing the Yagi-Yunes relation, and 
$\hat{a} = - 1.348 , \ \hat{b}= 0.357$ for our modified quasi-universal relation. 

\section{Conclusion}
\label{sec:conclusion}

We have tested if future GW detections might allow us to 
extract phenomenological relations between the spin-induced quadrupole moment 
and the tidal deformability of individual neutron stars. 
For this purpose, we have studied a simulated population of $120$ BNS systems for 
a 2G detector network and a 3G detector network.

We find that at design sensitivity a reduction of $50\%$ in the quadrupole moment would be visible, 
we anticipate that smaller deviations might not be observable.
However, this means that Advanced LIGO and Advanced Virgo might be able to detect 
possible deviations from existing, theoretically-predicted, quasi-universal relations. 
However, one would need a 3G detector network for a more reliable measurement. 
We find that with a network of 2 Cosmic Explorer-like detectors and 1 Einstein Telescope, 
we would be able to extract quasi-universal relations from the neutron star properties inferred from 
the analysis of the gravitational wave signals.

In the hypothetical scenario in which the extracted quasi-universal relations are not in agreement 
with theoretical predictions, this would either indicate a violation 
of general relativity or that our current description of the interior 
of neutron stars is insufficient. 

\begin{acknowledgments}
  We thank Sebastian Khan and the LIGO-Virgo collaborations'  
  extreme matter group for helpful discussions. We 
  thank N.~V.~Krishnendu and Nathan K.~Johnson-McDaniel 
  for helpful feedback on the draft and going 
  through it carefully. 
  We also thank Nathan K.~Johnson-McDaniel and An Chen for 
  support setting up the 3G injections.
  AS and TD are supported by the research programme 
  of the Netherlands Organisation for Scientific Research (NWO).
  TD acknowledges support by the European Union’s Horizon 
  2020 research and innovation program under grant
  agreement No 749145, BNSmergers. 
  The authors are grateful for computational resources provided by the 
  LIGO Laboratory and supported by the National Science Foundation Grants PHY-0757058 and PHY-0823459.
%   We are grateful for computational 
%   resources provided by the Caltech cluster.
\end{acknowledgments}

\bibliography{bns_pe_qm}

\begin{thebibliography}{43}%
\makeatletter
\providecommand \@ifxundefined [1]{%
 \@ifx{#1\undefined}
}%
\providecommand \@ifnum [1]{%
 \ifnum #1\expandafter \@firstoftwo
 \else \expandafter \@secondoftwo
 \fi
}%
\providecommand \@ifx [1]{%
 \ifx #1\expandafter \@firstoftwo
 \else \expandafter \@secondoftwo
 \fi
}%
\providecommand \natexlab [1]{#1}%
\providecommand \enquote  [1]{``#1''}%
\providecommand \bibnamefont  [1]{#1}%
\providecommand \bibfnamefont [1]{#1}%
\providecommand \citenamefont [1]{#1}%
\providecommand \href@noop [0]{\@secondoftwo}%
\providecommand \href [0]{\begingroup \@sanitize@url \@href}%
\providecommand \@href[1]{\@@startlink{#1}\@@href}%
\providecommand \@@href[1]{\endgroup#1\@@endlink}%
\providecommand \@sanitize@url [0]{\catcode `\\12\catcode `\$12\catcode
  `\&12\catcode `\#12\catcode `\^12\catcode `\_12\catcode `\%12\relax}%
\providecommand \@@startlink[1]{}%
\providecommand \@@endlink[0]{}%
\providecommand \url  [0]{\begingroup\@sanitize@url \@url }%
\providecommand \@url [1]{\endgroup\@href {#1}{\urlprefix }}%
\providecommand \urlprefix  [0]{URL }%
\providecommand \Eprint [0]{\href }%
\providecommand \doibase [0]{http://dx.doi.org/}%
\providecommand \selectlanguage [0]{\@gobble}%
\providecommand \bibinfo  [0]{\@secondoftwo}%
\providecommand \bibfield  [0]{\@secondoftwo}%
\providecommand \translation [1]{[#1]}%
\providecommand \BibitemOpen [0]{}%
\providecommand \bibitemStop [0]{}%
\providecommand \bibitemNoStop [0]{.\EOS\space}%
\providecommand \EOS [0]{\spacefactor3000\relax}%
\providecommand \BibitemShut  [1]{\csname bibitem#1\endcsname}%
\let\auto@bib@innerbib\@empty
%</preamble>
\bibitem [{\citenamefont {Abbott}\ \emph {et~al.}(2017)\citenamefont {Abbott}
  \emph {et~al.}}]{TheLIGOScientific:2017qsa}%
  \BibitemOpen
  \bibfield  {author} {\bibinfo {author} {\bibfnamefont {B.~P.}\ \bibnamefont
  {Abbott}} \emph {et~al.} (\bibinfo {collaboration} {Virgo, LIGO
  Scientific}),\ }\href {\doibase 10.1103/PhysRevLett.119.161101} {\bibfield
  {journal} {\bibinfo  {journal} {Phys. Rev. Lett.}\ }\textbf {\bibinfo
  {volume} {119}},\ \bibinfo {pages} {161101} (\bibinfo {year} {2017})},\
  \Eprint {http://arxiv.org/abs/1710.05832} {arXiv:1710.05832 [gr-qc]}
  \BibitemShut {NoStop}%
%%CITATION = ARXIV:1710.05832;%%
\bibitem [{\citenamefont {Abbott}\ \emph {et~al.}(2018)\citenamefont {Abbott}
  \emph {et~al.}}]{Abbott:2018exr}%
  \BibitemOpen
  \bibfield  {author} {\bibinfo {author} {\bibfnamefont {B.~P.}\ \bibnamefont
  {Abbott}} \emph {et~al.} (\bibinfo {collaboration} {Virgo, LIGO
  Scientific}),\ }\href@noop {} {\  (\bibinfo {year} {2018})},\ \Eprint
  {http://arxiv.org/abs/1805.11581} {arXiv:1805.11581 [gr-qc]} \BibitemShut
  {NoStop}%
%%CITATION = ARXIV:1805.11581;%%
\bibitem [{\citenamefont {Abbott}\ \emph {et~al.}(2019)\citenamefont {Abbott}
  \emph {et~al.}}]{Abbott:2018wiz}%
  \BibitemOpen
  \bibfield  {author} {\bibinfo {author} {\bibfnamefont {B.~P.}\ \bibnamefont
  {Abbott}} \emph {et~al.} (\bibinfo {collaboration} {LIGO Scientific,
  Virgo}),\ }\href {\doibase 10.1103/PhysRevX.9.011001} {\bibfield  {journal}
  {\bibinfo  {journal} {Phys. Rev.}\ }\textbf {\bibinfo {volume} {X9}},\
  \bibinfo {pages} {011001} (\bibinfo {year} {2019})},\ \Eprint
  {http://arxiv.org/abs/1805.11579} {arXiv:1805.11579 [gr-qc]} \BibitemShut
  {NoStop}%
%%CITATION = ARXIV:1805.11579;%%
\bibitem [{\citenamefont {Annala}\ \emph {et~al.}(2018)\citenamefont {Annala},
  \citenamefont {Gorda}, \citenamefont {Kurkela},\ and\ \citenamefont
  {Vuorinen}}]{Annala:2017llu}%
  \BibitemOpen
  \bibfield  {author} {\bibinfo {author} {\bibfnamefont {E.}~\bibnamefont
  {Annala}}, \bibinfo {author} {\bibfnamefont {T.}~\bibnamefont {Gorda}},
  \bibinfo {author} {\bibfnamefont {A.}~\bibnamefont {Kurkela}}, \ and\
  \bibinfo {author} {\bibfnamefont {A.}~\bibnamefont {Vuorinen}},\ }\href
  {\doibase 10.1103/PhysRevLett.120.172703} {\bibfield  {journal} {\bibinfo
  {journal} {Phys. Rev. Lett.}\ }\textbf {\bibinfo {volume} {120}},\ \bibinfo
  {pages} {172703} (\bibinfo {year} {2018})},\ \Eprint
  {http://arxiv.org/abs/1711.02644} {arXiv:1711.02644 [astro-ph.HE]}
  \BibitemShut {NoStop}%
%%CITATION = ARXIV:1711.02644;%%
\bibitem [{\citenamefont {Capano}\ \emph {et~al.}(2019)\citenamefont {Capano},
  \citenamefont {Tews}, \citenamefont {Brown}, \citenamefont {Margalit},
  \citenamefont {De}, \citenamefont {Kumar}, \citenamefont {Brown},
  \citenamefont {Krishnan},\ and\ \citenamefont {Reddy}}]{Capano:2019eae}%
  \BibitemOpen
  \bibfield  {author} {\bibinfo {author} {\bibfnamefont {C.~D.}\ \bibnamefont
  {Capano}}, \bibinfo {author} {\bibfnamefont {I.}~\bibnamefont {Tews}},
  \bibinfo {author} {\bibfnamefont {S.~M.}\ \bibnamefont {Brown}}, \bibinfo
  {author} {\bibfnamefont {B.}~\bibnamefont {Margalit}}, \bibinfo {author}
  {\bibfnamefont {S.}~\bibnamefont {De}}, \bibinfo {author} {\bibfnamefont
  {S.}~\bibnamefont {Kumar}}, \bibinfo {author} {\bibfnamefont {D.~A.}\
  \bibnamefont {Brown}}, \bibinfo {author} {\bibfnamefont {B.}~\bibnamefont
  {Krishnan}}, \ and\ \bibinfo {author} {\bibfnamefont {S.}~\bibnamefont
  {Reddy}},\ }\href@noop {} {\  (\bibinfo {year} {2019})},\ \Eprint
  {http://arxiv.org/abs/1908.10352} {arXiv:1908.10352 [astro-ph.HE]}
  \BibitemShut {NoStop}%
%%CITATION = ARXIV:1908.10352;%%
\bibitem [{\citenamefont {Bauswein}\ \emph {et~al.}(2017)\citenamefont
  {Bauswein}, \citenamefont {Just}, \citenamefont {Janka},\ and\ \citenamefont
  {Stergioulas}}]{Bauswein:2017vtn}%
  \BibitemOpen
  \bibfield  {author} {\bibinfo {author} {\bibfnamefont {A.}~\bibnamefont
  {Bauswein}}, \bibinfo {author} {\bibfnamefont {O.}~\bibnamefont {Just}},
  \bibinfo {author} {\bibfnamefont {H.-T.}\ \bibnamefont {Janka}}, \ and\
  \bibinfo {author} {\bibfnamefont {N.}~\bibnamefont {Stergioulas}},\ }\href
  {\doibase 10.3847/2041-8213/aa9994} {\bibfield  {journal} {\bibinfo
  {journal} {Astrophys. J.}\ }\textbf {\bibinfo {volume} {850}},\ \bibinfo
  {pages} {L34} (\bibinfo {year} {2017})},\ \Eprint
  {http://arxiv.org/abs/1710.06843} {arXiv:1710.06843 [astro-ph.HE]}
  \BibitemShut {NoStop}%
%%CITATION = ARXIV:1710.06843;%%
\bibitem [{\citenamefont {De}\ \emph {et~al.}(2018)\citenamefont {De},
  \citenamefont {Finstad}, \citenamefont {Lattimer}, \citenamefont {Brown},
  \citenamefont {Berger},\ and\ \citenamefont {Biwer}}]{De:2018uhw}%
  \BibitemOpen
  \bibfield  {author} {\bibinfo {author} {\bibfnamefont {S.}~\bibnamefont
  {De}}, \bibinfo {author} {\bibfnamefont {D.}~\bibnamefont {Finstad}},
  \bibinfo {author} {\bibfnamefont {J.~M.}\ \bibnamefont {Lattimer}}, \bibinfo
  {author} {\bibfnamefont {D.~A.}\ \bibnamefont {Brown}}, \bibinfo {author}
  {\bibfnamefont {E.}~\bibnamefont {Berger}}, \ and\ \bibinfo {author}
  {\bibfnamefont {C.~M.}\ \bibnamefont {Biwer}},\ }\href@noop {} {\  (\bibinfo
  {year} {2018})},\ \Eprint {http://arxiv.org/abs/1804.08583} {arXiv:1804.08583
  [astro-ph.HE]} \BibitemShut {NoStop}%
%%CITATION = ARXIV:1804.08583;%%
\bibitem [{\citenamefont {Margalit}\ and\ \citenamefont
  {Metzger}(2017)}]{Margalit:2017dij}%
  \BibitemOpen
  \bibfield  {author} {\bibinfo {author} {\bibfnamefont {B.}~\bibnamefont
  {Margalit}}\ and\ \bibinfo {author} {\bibfnamefont {B.~D.}\ \bibnamefont
  {Metzger}},\ }\href {\doibase 10.3847/2041-8213/aa991c} {\bibfield  {journal}
  {\bibinfo  {journal} {Astrophys. J.}\ }\textbf {\bibinfo {volume} {850}},\
  \bibinfo {pages} {L19} (\bibinfo {year} {2017})},\ \Eprint
  {http://arxiv.org/abs/1710.05938} {arXiv:1710.05938 [astro-ph.HE]}
  \BibitemShut {NoStop}%
%%CITATION = ARXIV:1710.05938;%%
\bibitem [{\citenamefont {Most}\ \emph {et~al.}(2018)\citenamefont {Most},
  \citenamefont {Weih}, \citenamefont {Rezzolla},\ and\ \citenamefont
  {Schaffner-Bielich}}]{Most:2018hfd}%
  \BibitemOpen
  \bibfield  {author} {\bibinfo {author} {\bibfnamefont {E.~R.}\ \bibnamefont
  {Most}}, \bibinfo {author} {\bibfnamefont {L.~R.}\ \bibnamefont {Weih}},
  \bibinfo {author} {\bibfnamefont {L.}~\bibnamefont {Rezzolla}}, \ and\
  \bibinfo {author} {\bibfnamefont {J.}~\bibnamefont {Schaffner-Bielich}},\
  }\href@noop {} {\  (\bibinfo {year} {2018})},\ \Eprint
  {http://arxiv.org/abs/1803.00549} {arXiv:1803.00549 [gr-qc]} \BibitemShut
  {NoStop}%
%%CITATION = ARXIV:1803.00549;%%
\bibitem [{\citenamefont {Coughlin}\ \emph {et~al.}(2018)\citenamefont
  {Coughlin} \emph {et~al.}}]{Coughlin:2018miv}%
  \BibitemOpen
  \bibfield  {author} {\bibinfo {author} {\bibfnamefont {M.~W.}\ \bibnamefont
  {Coughlin}} \emph {et~al.},\ }\href@noop {} {\  (\bibinfo {year} {2018})},\
  \Eprint {http://arxiv.org/abs/1805.09371} {arXiv:1805.09371 [astro-ph.HE]}
  \BibitemShut {NoStop}%
%%CITATION = ARXIV:1805.09371;%%
\bibitem [{\citenamefont {Radice}\ and\ \citenamefont
  {Dai}(2018)}]{Radice:2018ozg}%
  \BibitemOpen
  \bibfield  {author} {\bibinfo {author} {\bibfnamefont {D.}~\bibnamefont
  {Radice}}\ and\ \bibinfo {author} {\bibfnamefont {L.}~\bibnamefont {Dai}},\
  }\href@noop {} {\  (\bibinfo {year} {2018})},\ \Eprint
  {http://arxiv.org/abs/1810.12917} {arXiv:1810.12917 [astro-ph.HE]}
  \BibitemShut {NoStop}%
%%CITATION = ARXIV:1810.12917;%%
\bibitem [{\citenamefont {Abbott}\ \emph {et~al.}(2020)\citenamefont {Abbott}
  \emph {et~al.}}]{Abbott:2020uma}%
  \BibitemOpen
  \bibfield  {author} {\bibinfo {author} {\bibfnamefont {B.~P.}\ \bibnamefont
  {Abbott}} \emph {et~al.} (\bibinfo {collaboration} {LIGO Scientific,
  Virgo}),\ }\href@noop {} {\  (\bibinfo {year} {2020})},\ \Eprint
  {http://arxiv.org/abs/2001.01761} {arXiv:2001.01761 [astro-ph.HE]}
  \BibitemShut {NoStop}%
%%CITATION = ARXIV:2001.01761;%%
\bibitem [{\citenamefont {Coughlin}\ \emph
  {et~al.}(2019{\natexlab{a}})\citenamefont {Coughlin} \emph
  {et~al.}}]{Coughlin:2019xfb}%
  \BibitemOpen
  \bibfield  {author} {\bibinfo {author} {\bibfnamefont {M.~W.}\ \bibnamefont
  {Coughlin}} \emph {et~al.},\ }\href {\doibase 10.3847/2041-8213/ab4ad8}
  {\bibfield  {journal} {\bibinfo  {journal} {Astrophys. J.}\ }\textbf
  {\bibinfo {volume} {885}},\ \bibinfo {pages} {L19} (\bibinfo {year}
  {2019}{\natexlab{a}})},\ \Eprint {http://arxiv.org/abs/1907.12645}
  {arXiv:1907.12645 [astro-ph.HE]} \BibitemShut {NoStop}%
%%CITATION = ARXIV:1907.12645;%%
\bibitem [{\citenamefont {Coughlin}\ \emph
  {et~al.}(2019{\natexlab{b}})\citenamefont {Coughlin}, \citenamefont
  {Dietrich}, \citenamefont {Antier}, \citenamefont {Bulla}, \citenamefont
  {Foucart}, \citenamefont {Hotokezaka}, \citenamefont {Raaijmakers},
  \citenamefont {Hinderer},\ and\ \citenamefont {Nissanke}}]{Coughlin:2019zqi}%
  \BibitemOpen
  \bibfield  {author} {\bibinfo {author} {\bibfnamefont {M.~W.}\ \bibnamefont
  {Coughlin}}, \bibinfo {author} {\bibfnamefont {T.}~\bibnamefont {Dietrich}},
  \bibinfo {author} {\bibfnamefont {S.}~\bibnamefont {Antier}}, \bibinfo
  {author} {\bibfnamefont {M.}~\bibnamefont {Bulla}}, \bibinfo {author}
  {\bibfnamefont {F.}~\bibnamefont {Foucart}}, \bibinfo {author} {\bibfnamefont
  {K.}~\bibnamefont {Hotokezaka}}, \bibinfo {author} {\bibfnamefont
  {G.}~\bibnamefont {Raaijmakers}}, \bibinfo {author} {\bibfnamefont
  {T.}~\bibnamefont {Hinderer}}, \ and\ \bibinfo {author} {\bibfnamefont
  {S.}~\bibnamefont {Nissanke}},\ }\href@noop {} {\  (\bibinfo {year}
  {2019}{\natexlab{b}})},\ \Eprint {http://arxiv.org/abs/1910.11246}
  {arXiv:1910.11246 [astro-ph.HE]} \BibitemShut {NoStop}%
%%CITATION = ARXIV:1910.11246;%%
\bibitem [{\citenamefont {Laarakkers}\ and\ \citenamefont
  {Poisson}(1999)}]{Laarakkers:1997hb}%
  \BibitemOpen
  \bibfield  {author} {\bibinfo {author} {\bibfnamefont {W.~G.}\ \bibnamefont
  {Laarakkers}}\ and\ \bibinfo {author} {\bibfnamefont {E.}~\bibnamefont
  {Poisson}},\ }\href {\doibase 10.1086/306732} {\bibfield  {journal} {\bibinfo
   {journal} {Astrophys. J.}\ }\textbf {\bibinfo {volume} {512}},\ \bibinfo
  {pages} {282} (\bibinfo {year} {1999})},\ \Eprint
  {http://arxiv.org/abs/gr-qc/9709033} {arXiv:gr-qc/9709033 [gr-qc]}
  \BibitemShut {NoStop}%
%%CITATION = GR-QC/9709033;%%
\bibitem [{\citenamefont {Pappas}\ and\ \citenamefont
  {Apostolatos}(2012)}]{Pappas:2012ns}%
  \BibitemOpen
  \bibfield  {author} {\bibinfo {author} {\bibfnamefont {G.}~\bibnamefont
  {Pappas}}\ and\ \bibinfo {author} {\bibfnamefont {T.~A.}\ \bibnamefont
  {Apostolatos}},\ }\href {\doibase 10.1103/PhysRevLett.108.231104} {\bibfield
  {journal} {\bibinfo  {journal} {Phys. Rev. Lett.}\ }\textbf {\bibinfo
  {volume} {108}},\ \bibinfo {pages} {231104} (\bibinfo {year} {2012})},\
  \Eprint {http://arxiv.org/abs/1201.6067} {arXiv:1201.6067 [gr-qc]}
  \BibitemShut {NoStop}%
%%CITATION = ARXIV:1201.6067;%%
\bibitem [{\citenamefont {Poisson}(1998)}]{Poisson:1997ha}%
  \BibitemOpen
  \bibfield  {author} {\bibinfo {author} {\bibfnamefont {E.}~\bibnamefont
  {Poisson}},\ }\href {\doibase 10.1103/PhysRevD.57.5287} {\bibfield  {journal}
  {\bibinfo  {journal} {Phys. Rev.}\ }\textbf {\bibinfo {volume} {D57}},\
  \bibinfo {pages} {5287} (\bibinfo {year} {1998})},\ \Eprint
  {http://arxiv.org/abs/gr-qc/9709032} {arXiv:gr-qc/9709032 [gr-qc]}
  \BibitemShut {NoStop}%
%%CITATION = GR-QC/9709032;%%
\bibitem [{\citenamefont {Harry}\ and\ \citenamefont
  {Hinderer}(2018)}]{Harry:2018hke}%
  \BibitemOpen
  \bibfield  {author} {\bibinfo {author} {\bibfnamefont {I.}~\bibnamefont
  {Harry}}\ and\ \bibinfo {author} {\bibfnamefont {T.}~\bibnamefont
  {Hinderer}},\ }\href {\doibase 10.1088/1361-6382/aac7e3} {\bibfield
  {journal} {\bibinfo  {journal} {Class. Quant. Grav.}\ }\textbf {\bibinfo
  {volume} {35}},\ \bibinfo {pages} {145010} (\bibinfo {year} {2018})},\
  \Eprint {http://arxiv.org/abs/1801.09972} {arXiv:1801.09972 [gr-qc]}
  \BibitemShut {NoStop}%
%%CITATION = ARXIV:1801.09972;%%
\bibitem [{\citenamefont {Samajdar}\ and\ \citenamefont
  {Dietrich}(2019)}]{Samajdar:2019ulq}%
  \BibitemOpen
  \bibfield  {author} {\bibinfo {author} {\bibfnamefont {A.}~\bibnamefont
  {Samajdar}}\ and\ \bibinfo {author} {\bibfnamefont {T.}~\bibnamefont
  {Dietrich}},\ }\href {\doibase 10.1103/PhysRevD.100.024046} {\bibfield
  {journal} {\bibinfo  {journal} {Phys. Rev.}\ }\textbf {\bibinfo {volume}
  {D100}},\ \bibinfo {pages} {024046} (\bibinfo {year} {2019})},\ \Eprint
  {http://arxiv.org/abs/1905.03118} {arXiv:1905.03118 [gr-qc]} \BibitemShut
  {NoStop}%
%%CITATION = ARXIV:1905.03118;%%
\bibitem [{\citenamefont {Agathos}\ \emph {et~al.}(2015)\citenamefont
  {Agathos}, \citenamefont {Meidam}, \citenamefont {Del~Pozzo}, \citenamefont
  {Li}, \citenamefont {Tompitak}, \citenamefont {Veitch}, \citenamefont
  {Vitale},\ and\ \citenamefont {Van Den~Broeck}}]{Agathos:2015uaa}%
  \BibitemOpen
  \bibfield  {author} {\bibinfo {author} {\bibfnamefont {M.}~\bibnamefont
  {Agathos}}, \bibinfo {author} {\bibfnamefont {J.}~\bibnamefont {Meidam}},
  \bibinfo {author} {\bibfnamefont {W.}~\bibnamefont {Del~Pozzo}}, \bibinfo
  {author} {\bibfnamefont {T.~G.~F.}\ \bibnamefont {Li}}, \bibinfo {author}
  {\bibfnamefont {M.}~\bibnamefont {Tompitak}}, \bibinfo {author}
  {\bibfnamefont {J.}~\bibnamefont {Veitch}}, \bibinfo {author} {\bibfnamefont
  {S.}~\bibnamefont {Vitale}}, \ and\ \bibinfo {author} {\bibfnamefont
  {C.}~\bibnamefont {Van Den~Broeck}},\ }\href {\doibase
  10.1103/PhysRevD.92.023012} {\bibfield  {journal} {\bibinfo  {journal} {Phys.
  Rev.}\ }\textbf {\bibinfo {volume} {D92}},\ \bibinfo {pages} {023012}
  (\bibinfo {year} {2015})},\ \Eprint {http://arxiv.org/abs/1503.05405}
  {arXiv:1503.05405 [gr-qc]} \BibitemShut {NoStop}%
%%CITATION = ARXIV:1503.05405;%%
\bibitem [{\citenamefont {Krishnendu}\ \emph {et~al.}(2017)\citenamefont
  {Krishnendu}, \citenamefont {Arun},\ and\ \citenamefont
  {Mishra}}]{Krishnendu:2017shb}%
  \BibitemOpen
  \bibfield  {author} {\bibinfo {author} {\bibfnamefont {N.~V.}\ \bibnamefont
  {Krishnendu}}, \bibinfo {author} {\bibfnamefont {K.~G.}\ \bibnamefont
  {Arun}}, \ and\ \bibinfo {author} {\bibfnamefont {C.~K.}\ \bibnamefont
  {Mishra}},\ }\href {\doibase 10.1103/PhysRevLett.119.091101} {\bibfield
  {journal} {\bibinfo  {journal} {Phys. Rev. Lett.}\ }\textbf {\bibinfo
  {volume} {119}},\ \bibinfo {pages} {091101} (\bibinfo {year} {2017})},\
  \Eprint {http://arxiv.org/abs/1701.06318} {arXiv:1701.06318 [gr-qc]}
  \BibitemShut {NoStop}%
%%CITATION = ARXIV:1701.06318;%%
\bibitem [{\citenamefont {Krishnendu}\ \emph {et~al.}(2019)\citenamefont
  {Krishnendu}, \citenamefont {Saleem}, \citenamefont {Samajdar}, \citenamefont
  {Arun}, \citenamefont {Del~Pozzo},\ and\ \citenamefont
  {Mishra}}]{Krishnendu:2019tjp}%
  \BibitemOpen
  \bibfield  {author} {\bibinfo {author} {\bibfnamefont {N.~V.}\ \bibnamefont
  {Krishnendu}}, \bibinfo {author} {\bibfnamefont {M.}~\bibnamefont {Saleem}},
  \bibinfo {author} {\bibfnamefont {A.}~\bibnamefont {Samajdar}}, \bibinfo
  {author} {\bibfnamefont {K.~G.}\ \bibnamefont {Arun}}, \bibinfo {author}
  {\bibfnamefont {W.}~\bibnamefont {Del~Pozzo}}, \ and\ \bibinfo {author}
  {\bibfnamefont {C.~K.}\ \bibnamefont {Mishra}},\ }\href {\doibase
  10.1103/PhysRevD.100.104019} {\bibfield  {journal} {\bibinfo  {journal}
  {Phys. Rev.}\ }\textbf {\bibinfo {volume} {D100}},\ \bibinfo {pages} {104019}
  (\bibinfo {year} {2019})},\ \Eprint {http://arxiv.org/abs/1908.02247}
  {arXiv:1908.02247 [gr-qc]} \BibitemShut {NoStop}%
%%CITATION = ARXIV:1908.02247;%%
\bibitem [{\citenamefont {Yagi}\ and\ \citenamefont
  {Yunes}(2013{\natexlab{a}})}]{Yagi:2013bca}%
  \BibitemOpen
  \bibfield  {author} {\bibinfo {author} {\bibfnamefont {K.}~\bibnamefont
  {Yagi}}\ and\ \bibinfo {author} {\bibfnamefont {N.}~\bibnamefont {Yunes}},\
  }\href {\doibase 10.1126/science.1236462} {\bibfield  {journal} {\bibinfo
  {journal} {Science}\ }\textbf {\bibinfo {volume} {341}},\ \bibinfo {pages}
  {365} (\bibinfo {year} {2013}{\natexlab{a}})},\ \Eprint
  {http://arxiv.org/abs/1302.4499} {arXiv:1302.4499 [gr-qc]} \BibitemShut
  {NoStop}%
%%CITATION = ARXIV:1302.4499;%%
\bibitem [{\citenamefont {Carson}\ \emph {et~al.}(2019)\citenamefont {Carson},
  \citenamefont {Chatziioannou}, \citenamefont {Haster}, \citenamefont {Yagi},\
  and\ \citenamefont {Yunes}}]{Carson:2019rjx}%
  \BibitemOpen
  \bibfield  {author} {\bibinfo {author} {\bibfnamefont {Z.}~\bibnamefont
  {Carson}}, \bibinfo {author} {\bibfnamefont {K.}~\bibnamefont
  {Chatziioannou}}, \bibinfo {author} {\bibfnamefont {C.-J.}\ \bibnamefont
  {Haster}}, \bibinfo {author} {\bibfnamefont {K.}~\bibnamefont {Yagi}}, \ and\
  \bibinfo {author} {\bibfnamefont {N.}~\bibnamefont {Yunes}},\ }\href
  {\doibase 10.1103/PhysRevD.99.083016} {\bibfield  {journal} {\bibinfo
  {journal} {Phys. Rev.}\ }\textbf {\bibinfo {volume} {D99}},\ \bibinfo {pages}
  {083016} (\bibinfo {year} {2019})},\ \Eprint
  {http://arxiv.org/abs/1903.03909} {arXiv:1903.03909 [gr-qc]} \BibitemShut
  {NoStop}%
%%CITATION = ARXIV:1903.03909;%%
\bibitem [{\citenamefont {Yagi}\ and\ \citenamefont
  {Yunes}(2013{\natexlab{b}})}]{Yagi:2013awa}%
  \BibitemOpen
  \bibfield  {author} {\bibinfo {author} {\bibfnamefont {K.}~\bibnamefont
  {Yagi}}\ and\ \bibinfo {author} {\bibfnamefont {N.}~\bibnamefont {Yunes}},\
  }\href {\doibase 10.1103/PhysRevD.88.023009} {\bibfield  {journal} {\bibinfo
  {journal} {Phys. Rev.}\ }\textbf {\bibinfo {volume} {D88}},\ \bibinfo {pages}
  {023009} (\bibinfo {year} {2013}{\natexlab{b}})},\ \Eprint
  {http://arxiv.org/abs/1303.1528} {arXiv:1303.1528 [gr-qc]} \BibitemShut
  {NoStop}%
%%CITATION = ARXIV:1303.1528;%%
\bibitem [{\citenamefont {Veitch}\ \emph {et~al.}(2015)\citenamefont {Veitch}
  \emph {et~al.}}]{Veitch:2014wba}%
  \BibitemOpen
  \bibfield  {author} {\bibinfo {author} {\bibfnamefont {J.}~\bibnamefont
  {Veitch}} \emph {et~al.},\ }\href {\doibase 10.1103/PhysRevD.91.042003}
  {\bibfield  {journal} {\bibinfo  {journal} {Phys. Rev.}\ }\textbf {\bibinfo
  {volume} {D91}},\ \bibinfo {pages} {042003} (\bibinfo {year} {2015})},\
  \Eprint {http://arxiv.org/abs/1409.7215} {arXiv:1409.7215 [gr-qc]}
  \BibitemShut {NoStop}%
%%CITATION = ARXIV:1409.7215;%%
\bibitem [{\citenamefont {{LIGO Scientific Collaboration}}(2018)}]{lalsuite}%
  \BibitemOpen
  \bibfield  {author} {\bibinfo {author} {\bibnamefont {{LIGO Scientific
  Collaboration}}},\ }\href {\doibase 10.7935/GT1W-FZ16} {\enquote {\bibinfo
  {title} {{LIGO} {A}lgorithm {L}ibrary - {LALS}uite},}\ }\bibinfo
  {howpublished} {free software (GPL)} (\bibinfo {year} {2018})\BibitemShut
  {NoStop}%
\bibitem [{\citenamefont {Veitch}\ and\ \citenamefont
  {Vecchio}(2010)}]{Veitch:2009hd}%
  \BibitemOpen
  \bibfield  {author} {\bibinfo {author} {\bibfnamefont {J.}~\bibnamefont
  {Veitch}}\ and\ \bibinfo {author} {\bibfnamefont {A.}~\bibnamefont
  {Vecchio}},\ }\href {\doibase 10.1103/PhysRevD.81.062003} {\bibfield
  {journal} {\bibinfo  {journal} {Phys. Rev.}\ }\textbf {\bibinfo {volume}
  {D81}},\ \bibinfo {pages} {062003} (\bibinfo {year} {2010})},\ \Eprint
  {http://arxiv.org/abs/0911.3820} {arXiv:0911.3820 [astro-ph.CO]} \BibitemShut
  {NoStop}%
%%CITATION = ARXIV:0911.3820;%%
\bibitem [{\citenamefont {Skilling}(2006)}]{skilling2006}%
  \BibitemOpen
  \bibfield  {author} {\bibinfo {author} {\bibfnamefont {J.}~\bibnamefont
  {Skilling}},\ }\href {\doibase 10.1214/06-BA127} {\bibfield  {journal}
  {\bibinfo  {journal} {Bayesian Anal.}\ }\textbf {\bibinfo {volume} {1}},\
  \bibinfo {pages} {833} (\bibinfo {year} {2006})}\BibitemShut {NoStop}%
\bibitem [{\citenamefont {Dietrich}\ \emph {et~al.}(2019)\citenamefont
  {Dietrich}, \citenamefont {Samajdar}, \citenamefont {Khan}, \citenamefont
  {Johnson-McDaniel}, \citenamefont {Dudi},\ and\ \citenamefont
  {Tichy}}]{Dietrich:2019kaq}%
  \BibitemOpen
  \bibfield  {author} {\bibinfo {author} {\bibfnamefont {T.}~\bibnamefont
  {Dietrich}}, \bibinfo {author} {\bibfnamefont {A.}~\bibnamefont {Samajdar}},
  \bibinfo {author} {\bibfnamefont {S.}~\bibnamefont {Khan}}, \bibinfo {author}
  {\bibfnamefont {N.~K.}\ \bibnamefont {Johnson-McDaniel}}, \bibinfo {author}
  {\bibfnamefont {R.}~\bibnamefont {Dudi}}, \ and\ \bibinfo {author}
  {\bibfnamefont {W.}~\bibnamefont {Tichy}},\ }\href {\doibase
  10.1103/PhysRevD.100.044003} {\bibfield  {journal} {\bibinfo  {journal}
  {Phys. Rev.}\ }\textbf {\bibinfo {volume} {D100}},\ \bibinfo {pages} {044003}
  (\bibinfo {year} {2019})},\ \Eprint {http://arxiv.org/abs/1905.06011}
  {arXiv:1905.06011 [gr-qc]} \BibitemShut {NoStop}%
%%CITATION = ARXIV:1905.06011;%%
\bibitem [{\citenamefont {Alford}\ \emph {et~al.}(2005)\citenamefont {Alford},
  \citenamefont {Braby}, \citenamefont {Paris},\ and\ \citenamefont
  {Reddy}}]{Alford:2004pf}%
  \BibitemOpen
  \bibfield  {author} {\bibinfo {author} {\bibfnamefont {M.}~\bibnamefont
  {Alford}}, \bibinfo {author} {\bibfnamefont {M.}~\bibnamefont {Braby}},
  \bibinfo {author} {\bibfnamefont {M.~W.}\ \bibnamefont {Paris}}, \ and\
  \bibinfo {author} {\bibfnamefont {S.}~\bibnamefont {Reddy}},\ }\href
  {\doibase 10.1086/430902} {\bibfield  {journal} {\bibinfo  {journal}
  {Astrophys. J.}\ }\textbf {\bibinfo {volume} {629}},\ \bibinfo {pages} {969}
  (\bibinfo {year} {2005})},\ \Eprint {http://arxiv.org/abs/nucl-th/0411016}
  {arXiv:nucl-th/0411016 [nucl-th]} \BibitemShut {NoStop}%
%%CITATION = NUCL-TH/0411016;%%
\bibitem [{\citenamefont {Akmal}\ \emph {et~al.}(1998)\citenamefont {Akmal},
  \citenamefont {Pandharipande},\ and\ \citenamefont
  {Ravenhall}}]{Akmal:1998cf}%
  \BibitemOpen
  \bibfield  {author} {\bibinfo {author} {\bibfnamefont {A.}~\bibnamefont
  {Akmal}}, \bibinfo {author} {\bibfnamefont {V.~R.}\ \bibnamefont
  {Pandharipande}}, \ and\ \bibinfo {author} {\bibfnamefont {D.~G.}\
  \bibnamefont {Ravenhall}},\ }\href {\doibase 10.1103/PhysRevC.58.1804}
  {\bibfield  {journal} {\bibinfo  {journal} {Phys. Rev.}\ }\textbf {\bibinfo
  {volume} {C58}},\ \bibinfo {pages} {1804} (\bibinfo {year} {1998})},\ \Eprint
  {http://arxiv.org/abs/nucl-th/9804027} {arXiv:nucl-th/9804027 [nucl-th]}
  \BibitemShut {NoStop}%
%%CITATION = NUCL-TH/9804027;%%
\bibitem [{\citenamefont {Lorimer}(2008)}]{Lorimer:2008se}%
  \BibitemOpen
  \bibfield  {author} {\bibinfo {author} {\bibfnamefont {D.~R.}\ \bibnamefont
  {Lorimer}},\ }\href {\doibase 10.12942/lrr-2008-8} {\bibfield  {journal}
  {\bibinfo  {journal} {Living Rev. Rel.}\ }\textbf {\bibinfo {volume} {11}},\
  \bibinfo {pages} {8} (\bibinfo {year} {2008})},\ \Eprint
  {http://arxiv.org/abs/0811.0762} {arXiv:0811.0762 [astro-ph]} \BibitemShut
  {NoStop}%
%%CITATION = ARXIV:0811.0762;%%
\bibitem [{\citenamefont {Lattimer}(2012)}]{Lattimer:2012nd}%
  \BibitemOpen
  \bibfield  {author} {\bibinfo {author} {\bibfnamefont {J.~M.}\ \bibnamefont
  {Lattimer}},\ }\href {\doibase 10.1146/annurev-nucl-102711-095018} {\bibfield
   {journal} {\bibinfo  {journal} {Ann. Rev. Nucl. Part. Sci.}\ }\textbf
  {\bibinfo {volume} {62}},\ \bibinfo {pages} {485} (\bibinfo {year} {2012})},\
  \Eprint {http://arxiv.org/abs/1305.3510} {arXiv:1305.3510 [nucl-th]}
  \BibitemShut {NoStop}%
%%CITATION = ARXIV:1305.3510;%%
\bibitem [{\citenamefont {Alexander}\ and\ \citenamefont
  {Yunes}(2009)}]{Alexander:2009tp}%
  \BibitemOpen
  \bibfield  {author} {\bibinfo {author} {\bibfnamefont {S.}~\bibnamefont
  {Alexander}}\ and\ \bibinfo {author} {\bibfnamefont {N.}~\bibnamefont
  {Yunes}},\ }\href {\doibase 10.1016/j.physrep.2009.07.002} {\bibfield
  {journal} {\bibinfo  {journal} {Phys. Rept.}\ }\textbf {\bibinfo {volume}
  {480}},\ \bibinfo {pages} {1} (\bibinfo {year} {2009})},\ \Eprint
  {http://arxiv.org/abs/0907.2562} {arXiv:0907.2562 [hep-th]} \BibitemShut
  {NoStop}%
%%CITATION = ARXIV:0907.2562;%%
\bibitem [{\citenamefont {Flanagan}\ and\ \citenamefont
  {Hinderer}(2008)}]{Flanagan:2007ix}%
  \BibitemOpen
  \bibfield  {author} {\bibinfo {author} {\bibfnamefont {E.~E.}\ \bibnamefont
  {Flanagan}}\ and\ \bibinfo {author} {\bibfnamefont {T.}~\bibnamefont
  {Hinderer}},\ }\href {\doibase 10.1103/PhysRevD.77.021502} {\bibfield
  {journal} {\bibinfo  {journal} {Phys. Rev.}\ }\textbf {\bibinfo {volume}
  {D77}},\ \bibinfo {pages} {021502} (\bibinfo {year} {2008})},\ \Eprint
  {http://arxiv.org/abs/0709.1915} {arXiv:0709.1915 [astro-ph]} \BibitemShut
  {NoStop}%
%%CITATION = ARXIV:0709.1915;%%
\bibitem [{Note1()}]{Note1}%
  \BibitemOpen
  \bibinfo {note} {We expect that this observation does not depend on the
  particular EOS, but that only the deviation from Eq.~\protect \textup {\hbox
  {\mathsurround \z@ \protect \normalfont (\ignorespaces \ref {eqn:qu}\unskip
  \@@italiccorr )}} is important}\BibitemShut {NoStop}%
\bibitem [{\citenamefont {Hild}\ \emph {et~al.}(2011)\citenamefont {Hild} \emph
  {et~al.}}]{Hild:2010id}%
  \BibitemOpen
  \bibfield  {author} {\bibinfo {author} {\bibfnamefont {S.}~\bibnamefont
  {Hild}} \emph {et~al.},\ }\href {\doibase 10.1088/0264-9381/28/9/094013}
  {\bibfield  {journal} {\bibinfo  {journal} {Class. Quant. Grav.}\ }\textbf
  {\bibinfo {volume} {28}},\ \bibinfo {pages} {094013} (\bibinfo {year}
  {2011})},\ \Eprint {http://arxiv.org/abs/1012.0908} {arXiv:1012.0908 [gr-qc]}
  \BibitemShut {NoStop}%
%%CITATION = ARXIV:1012.0908;%%
\bibitem [{\citenamefont {Punturo}\ \emph {et~al.}(2010)\citenamefont {Punturo}
  \emph {et~al.}}]{Punturo_2010}%
  \BibitemOpen
  \bibfield  {author} {\bibinfo {author} {\bibfnamefont {M.}~\bibnamefont
  {Punturo}} \emph {et~al.},\ }\href {\doibase 10.1088/0264-9381/27/19/194002}
  {\bibfield  {journal} {\bibinfo  {journal} {Classical and Quantum Gravity}\
  }\textbf {\bibinfo {volume} {27}},\ \bibinfo {pages} {194002} (\bibinfo
  {year} {2010})}\BibitemShut {NoStop}%
\bibitem [{\citenamefont {Dwyer}\ \emph {et~al.}(2015)\citenamefont {Dwyer},
  \citenamefont {Sigg}, \citenamefont {Ballmer}, \citenamefont {Barsotti},
  \citenamefont {Mavalvala},\ and\ \citenamefont {Evans}}]{Dwyer:2014fpa}%
  \BibitemOpen
  \bibfield  {author} {\bibinfo {author} {\bibfnamefont {S.}~\bibnamefont
  {Dwyer}}, \bibinfo {author} {\bibfnamefont {D.}~\bibnamefont {Sigg}},
  \bibinfo {author} {\bibfnamefont {S.~W.}\ \bibnamefont {Ballmer}}, \bibinfo
  {author} {\bibfnamefont {L.}~\bibnamefont {Barsotti}}, \bibinfo {author}
  {\bibfnamefont {N.}~\bibnamefont {Mavalvala}}, \ and\ \bibinfo {author}
  {\bibfnamefont {M.}~\bibnamefont {Evans}},\ }\href {\doibase
  10.1103/PhysRevD.91.082001} {\bibfield  {journal} {\bibinfo  {journal} {Phys.
  Rev.}\ }\textbf {\bibinfo {volume} {D91}},\ \bibinfo {pages} {082001}
  (\bibinfo {year} {2015})},\ \Eprint {http://arxiv.org/abs/1410.0612}
  {arXiv:1410.0612 [astro-ph.IM]} \BibitemShut {NoStop}%
%%CITATION = ARXIV:1410.0612;%%
\bibitem [{Note2()}]{Note2}%
  \BibitemOpen
  \bibinfo {note} {We note that we include an older CE noise curve and that we
  do not take a frequency-dependent response into account, we expect that due
  to our relatively large initial frequency, the effect of the
  frequency-dependence of the response function is small. We refer the reader
  to~\cite {Chen:2020fzm} for additional details.}\BibitemShut {Stop}%
\bibitem [{Note3()}]{Note3}%
  \BibitemOpen
  \bibinfo {note} {Although we use a $f_{\protect \rm low} =28$ Hz for the 3G
  network, we obtain SNR values of about $10^2$ up to $10^3$, i.e., about $20$
  to $30$ times larger than for the 2G network.}\BibitemShut {Stop}%
\bibitem [{\citenamefont {Chen}\ \emph {et~al.}(2020)\citenamefont {Chen},
  \citenamefont {Johnson-McDaniel}, \citenamefont {Dietrich},\ and\
  \citenamefont {Dudi}}]{Chen:2020fzm}%
  \BibitemOpen
  \bibfield  {author} {\bibinfo {author} {\bibfnamefont {A.}~\bibnamefont
  {Chen}}, \bibinfo {author} {\bibfnamefont {N.~K.}\ \bibnamefont
  {Johnson-McDaniel}}, \bibinfo {author} {\bibfnamefont {T.}~\bibnamefont
  {Dietrich}}, \ and\ \bibinfo {author} {\bibfnamefont {R.}~\bibnamefont
  {Dudi}},\ }\href@noop {} {\  (\bibinfo {year} {2020})},\ \Eprint
  {http://arxiv.org/abs/2001.11470} {arXiv:2001.11470 [astro-ph.HE]}
  \BibitemShut {NoStop}%
%%CITATION = ARXIV:2001.11470;%%
\end{thebibliography}%

\end{document}